\documentclass{emulateapj}

\newcommand{\gcc}{\ \mathrm{g\ cm^{-3} }}
\newcommand{\nuclei}[2]{\ensuremath{\mathrm{^{#1}#2}}}
\newcommand{\ye}{\ensuremath{Y_e }}

\begin{document} 
\title{Proton-rich Nuclear Statistical Equilibrium}

\author{I. R. Seitenzahl\altaffilmark{1,2,3}, F. X. Timmes\altaffilmark{2,4}, A. Marin-Lafl\`eche\altaffilmark{5,6},  E. Brown\altaffilmark{2,7}, G. Magkotsios\altaffilmark{2,8}, J. Truran\altaffilmark{2,3,5,9} } 
\altaffiltext{1}{Department of Physics,
                 The University of Chicago,
                 Chicago, IL  60637}
\altaffiltext{2}{Joint Institute for Nuclear Astrophysics (JINA)}
\altaffiltext{3}{Enrico Fermi Institute,
                 The University of Chicago,
                 Chicago, IL  60637}
\altaffiltext{4}{Department of Physics and Astronomy,
                 Arizona State University,
                 Tempe, AZ  85287}
\altaffiltext{5}{Department of Astronomy and Astrophysics,
                 The University of Chicago,
                 Chicago, IL  60637}
\altaffiltext{6}{Department of Physics and Astronomy,
                 \'Ecole Polytechnique,
                 Palaiseau, France}                 
\altaffiltext{7}{Department of Physics and Astronomy,
                 Michigan State University,
                 East Lansing, MI 48824}
\altaffiltext{8}{Department of Physics,
                 University of Notre Dame,           
                  Notre Dame, IN 46556}
\altaffiltext{9}{Argonne National Laboratory,
                         Argonne, IL 60439}
                 
\begin{abstract}
Proton-rich material in a state of nuclear statistical equilibrium (NSE) is one of the least studied regimes of nucleosynthesis. One reason for this is that after hydrogen burning, stellar evolution proceeds at conditions of equal number of neutrons and protons or at a slight degree of neutron-richness. Proton-rich nucleosynthesis in stars tends to occur only when hydrogen-rich material that accretes onto a white dwarf of neutron star explodes, or when neutrino interactions in the winds from a nascent proto-neutron star or collapsar-disk drive the matter proton-rich prior to or during the nucleosynthesis. In this paper we solve the NSE equations for a range of proton-rich thermodynamic conditions. We show that cold proton-rich NSE is qualitatively different from neutron-rich NSE. Instead of being dominated by the Fe-peak nuclei with the largest binding energy per nucleon that have a proton to nucleon ratio close to the prescribed electron fraction, NSE for proton-rich material near freeze-out temperature is mainly composed of \nuclei{56}{Ni} and free protons. Previous results of nuclear reaction network calculations rely on this non-intuitive high proton abundance, which this paper will explain. We show how the differences and especially the large fraction of free protons arises from the minimization of the free energy as a result of a delicate competition between the entropy and nuclear binding energy.  
\end{abstract}

\section{Introduction}
\label{sec:intro}
Recently, a new nucleosynthesis process was invented to explain the production of the proton-rich isotopes \nuclei{92,94}{Mo} and \nuclei{96,98}{Ru} \citep{frohlich06}. In this so called $\nu p$-process, matter ejected from the surface layers of a proto-neutron star is exposed to a strong neutrino flux which results in the following weak interactions: 
\begin{eqnarray}
\nu_e + n  &\leftrightharpoons& p + e^- \label{eq:one} \\ 
\bar{\nu}_e + p  &\leftrightharpoons& n + e^+ \label{eq:two}
\end{eqnarray}
The mass difference between the neutron and the proton causes the neutrino interactions to be dominated by the forward reaction of eq.~(\ref{eq:one}), which drives the nuclear matter proton-rich. The matter then expands and assembles into NSE, which is not only proton-rich (i.e. $Y_e > 0.5$) but actually contains a large mass fraction of free protons. 
The forward reaction of eq.~(\ref{eq:two}) converts some of the free protons to neutrons, which allows nuclear matter to move past the ``bottle neck" nucleus $\nuclei{64}{Ge}$ via the fast $\nuclei{64}{Ge}(n,p)\nuclei{64}{Ga}$ reaction and the aforementioned proton-rich isotopes are synthesized during the freeze out phase from the proton-rich NSE state \citep{frohlich06}. 
\citet{meyer_1994_aa}, \citet{jordan_2004_aa}, \citet{pruet05,pruet06} and \citet{wanajo06} calculate nucleosynthesis in similar environments. 
 
 We show that the NSE mass fractions exhibit a great degree of symmetry across the line $Y_e=0.5$ for high temperatures where the composition is dominated by free nucleons and $\nuclei{4}{He}.$ 
For colder temperatures, ($T_9 \leq 6.0$ and $\rho \sim 10^7 \gcc$), there are hardly any free neutrons present in the NSE state for $0.4 < \ye \leq 0.5$. 
A na\"ive guess based on symmetry of NSE abundances would therefore not lead one to expect many free protons during freeze out for $\ye > 0.5$ either. 
A large number of free protons is however observed to occur in nuclear reaction networks calculations.
 We show that for a relatively cold, near freeze-out temperature NSE state, there is indeed a qualitative difference in the mass fraction trends for $\ye > 0.5$ and $\ye < 0.5$. 
 Restricting ourselves to the Ni isotopic chain and free nucleons, we explain how this difference arises as a result of a competition between the temperature dependent entropy term and the nuclear binding energy term contribution to the free energy. 
 
\section{Nuclear Statistical Equilibrium equations}
\label{sec:nse}
NSE is established if all fusion reactions are in equilibrium with their inverses for a set of thermodynamic state variables and \ye\ \citep[e.g.][]{clifford65}. 
Detailed balance relates the chemical potential of a nucleus $^{Z_i+N_i}_{\ \ \ Z_i}X_i$ to those of free nucleons:
\mbox{$\mu_i = Z_i \mu_p + N_i \mu_n$}. This yields an expression for the number density of nucleus~$i$:
$n_i = g_i \Big(\frac{2\pi m_i kT}{h^2}\Big)^{3/2} \! \! \! \exp{\!\Big[\frac{Z_i (\mu^{kin}_p + \mu^{C}_p)+N_i \mu^{kin}_n -\mu^{C}_i+ Q_i}{kT}\Big]}.$
The NSE constraint equations can be perhaps most naturally written in a number density basis as $\sum_i A_i n_i = n_B$ and $\sum_i Z_i n_i=n_B Y_e$, where $n_B=n_n+n_p$ and $Y_e=n_p/n_n$. 
We solve the NSE constraint equations numerically for the kinetic chemical potentials $\mu^{kin}_p$ and $\mu^{kin}_n$ of the (assumed Maxwellian) protons and neutrons respectively.
For the Coulomb contributions $\mu^{C}_p$ and $\mu^{C}_i$ and the nuclear partition functions $g_i$ we use the same formalism as \citet{calder07} and \citet{seitenzahl08}. 
\section{Results}
\begin{figure}
\plotone{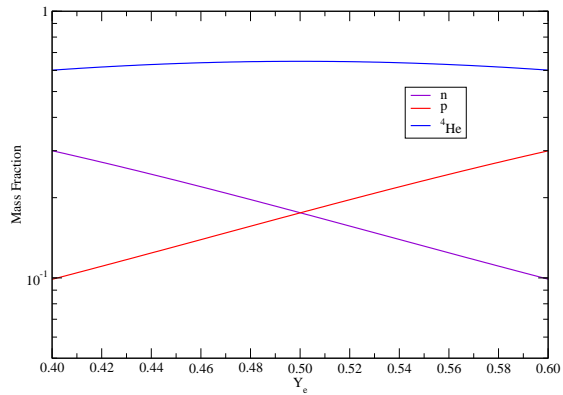}
\caption{\label{fig:highT} Mass fractions of nuclei in NSE as a function of electron fraction for a constant density $\rho = 10^7$ g cm$^{-3}$  \mbox{($n_B = 6.0\times10^{30}\,\mathrm{cm}^{-3}$)} and relatively high  temperature \mbox{$T = 9.0\times10^9\,\mathrm{K}$.} Shown are all nuclei with mass fractions larger than $10^{-2}$. The mass fractions exhibit large degree of (in the case of nucleons complementary) symmetry across $Y_e=0.5$.}
\end{figure}

The results presented here are all for a baryonic mass density of $\rho = 10^7$ g cm$^{-3}$, which corresponds
to a baryon number density of \mbox{$ n_B \approx 6.0\times10^{30}\,\mathrm{cm}^{-3}$}.
At high temperature ($T_9=9.0$) the NSE mass fractions are dominated by free nucleons and \nuclei{4}{He}. The \nuclei{4}{He} mass fraction is symmetric across the line $Y_e=0.5$, and the mass fractions of free protons and neutrons are symmetric in a complementary sense -- free protons are more abundant for $Y_e > 0.5$ and free neutrons are more abundant for $Y_e < 0.5$ (see fig.~\ref{fig:highT}).

\begin{figure}
\plotone{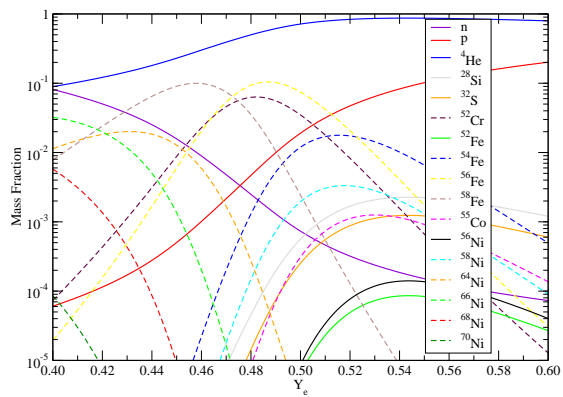}
\caption{\label{fig:midT} Mass fractions of nuclei in NSE as a function of electron fraction for a constant density $\rho = 10^7$ g cm$^{-3}$  \mbox{($n_B = 6.0\times10^{30}\,\mathrm{cm}^{-3}$)} and  temperature \mbox{$T = 6.5\times10^9\,\mathrm{K}$.} Shown are some abundant nuclei with mass fractions larger than $10^{-5}$. The symmetry across $Y_e=0.5$ is already broken.}
\end{figure}

At somewhat lower temperature ($T_9=6.5$), the symmetry of the NSE mass fractions across the line of self-conjugacy is broken and only qualitatively discernible. For $\Delta Y_e > 0$, free neutrons are less abundant for \mbox{$Y_e=0.5-\Delta Y_e$} than are free protons for \mbox{$Y_e=0.5+\Delta Y_e$} (see fig.~\ref{fig:midT}). Furthermore, the mass fraction of $\nuclei{4}{He}$ is not symmetric anymore and the abundance peaks of the Fe-peak nuclei are wider on the proton-rich side.

\begin{figure}
\plotone{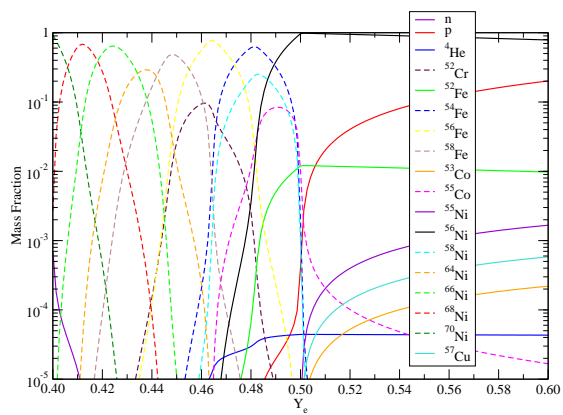}
\caption{\label{fig:lowT} Mass fractions of nuclei in NSE as a function of electron fraction for a constant density $\rho = 10^7$ g cm$^{-3}$  \mbox{($n_B = 6.0\times10^{30}\,\mathrm{cm}^{-3}$)} and temperature \mbox{$T = 3.5\times10^9\,\mathrm{K}$.} Shown are some abundant nuclei with mass fractions larger than $10^{-5}$. The mass fractions on either side of $Y_e=0.5$ exhibit qualitatively very different behavior.}
\end{figure}

At even lower temperature ($T_9=3.5$), the qualitative features of the mass fractions of nuclei in NSE as a function of $Y_e$ at fixed density and temperature change dramatically in the transition from the neutron-rich to the proton-rich side. 
For $Y_e < 0.5$, the mass fraction landscape is comprised of a sequence of overlapping abundance peaks \citep[cf. e.g.][]{clifford65,hartmann85,nadyozhin04}. 
Fe-peak nuclei with a proton to nucleon ratio equal or close to the prescribed $Y_e$ of the ensemble and a large binding energy per nucleon, $q/A$, are the most abundant nuclei. For $\ye > 0.5$, the picture changes abruptly (see fig. \ref{fig:lowT}). The mass fraction distributions of the Fe-peak nuclei are no longer peaked, but rather either slowly rising or falling.
\nuclei{56}{Ni} remains the most abundant nuclear species by mass all the way out past $\ye = 0.6$. 
The mass fraction of free protons continues to rise.  \nuclei{52}{Fe} shows a similar trend like \nuclei{56}{Ni}, albeit at a smaller abundance level. 
There is only a slow rise in the mass fractions of proton-rich Fe-peak nuclei for increasing \ye, the abundance peaks for the Fe-peak nuclei with a proton to nucleon ratio equal to $Y_e$ that are so prominent on the neutron-rich side are absent. 
Various animations of NSE in proton-rich environments may be downloaded from \url{http://cococubed.asu.edu/code\_pages/nse.shtml}.

\section{Discussion}
\label{sec:discussion}
In NSE, the Helmholtz free energy ${\mathcal F} = (U - Q ) -TS$ is minimized with respect to the nuclide mass fractions \citep[e.g.][]{nadyozhin05}.
Before we discuss and compare the terms that make up ${\mathcal F}$ for different choices of compositions, it is instructive to review the dependence of $q/A$ on the number of nucleons along an isotopic chain, the attributes of which ultimately are responsible for the lack of symmetry. 
\subsection{The nuclear binding energy}
\label{subsec:bindingenergy}
\begin{figure}
\plotone{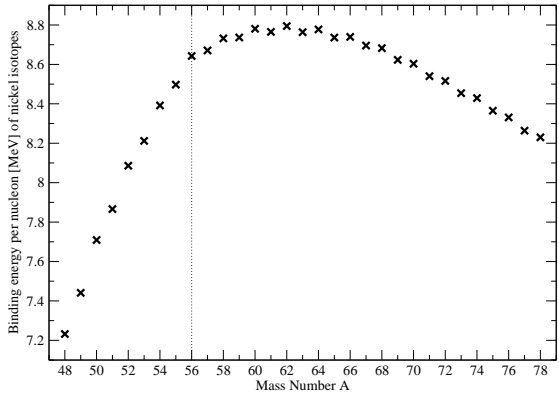}
\caption{\label{fig:BE-Ni} $q/A$ for Ni isotopes. Note the rapid decline on the proton-rich side of \nuclei{56}{Ni} and the particularly large gap between \nuclei{56}{Ni} and \nuclei{55}{Ni}.}
\end{figure}
$q/A$ on the proton-rich side for the Ni isotopic chain decreases rapidly towards proton drip, whereas it gently increases to a maximum at $A_{max}=62$ before slowly falling off towards neutron drip (see fig.~\ref{fig:BE-Ni}). 
Other isotopic chains, such as the one for Fe, look qualitatively very similar.
The shape of the $q/A$-curve can be qualitatively understood by considering a simple liquid-drop mass formula taking the volume, surface, asymmetry, and Coulomb terms into account
 \begin{equation}
 \label{eq:massformula}
 Q = a_{V} - a_{S}A^{-1/3} - a_{A}(1-2Z/A)^{2} - a_{C}Z^{2} A^{-4/3}
 \end{equation}
 For simplicity, we have neglected other terms such as pairing, Wigner, or residual interaction. The derivative with respect to nucleon number is given by
 \begin{equation}\label{eq:slope}
{\mathcal S}(A) = \left.\frac{\partial Q}{\partial A}\right|_{Z} = \frac{a_{S}}{3}A^{-4/3} +\frac{4a_{A}Z}{A}(1-2Z/A)+\frac{4a_{C}Z^2}{3}A^{-7/3}.
\end{equation}
 The second (asymmetry) term is positive on the proton-rich side, zero for $2Z=A$, and negative on the neutron-rich side, which gives the $q/A$-curve its general concave up shape. 
 It is, however, not entirely symmetric about the line $2Z=A$. 
 In fact, since
 \begin{equation}
  \Big|\frac{Z}{2Z+\Delta A}\big(1-\frac{2Z}{2Z+\Delta A}\big)\Big| < \Big|\frac{Z}{2Z-\Delta A}\big(1-\frac{2Z}{2Z-\Delta A}\big)\Big| 
  \end{equation}
  the absolute value of the slope due to the asymmetry term alone is larger on the proton-rich side. 
 The first (surface) and third (Coulomb) term are always positive, both monotonically increasing with $A$, 
 which consequently shifts the most tightly bound isotope to the neutron-rich side and further increases the asymmetry in the slopes on either side of the maximum.
 This simple argument shows the essence of how the surface and Coulomb terms in the nuclear mass formula leads to a more rapid fall off in $q/A$ on the proton-rich side. 
\nuclei{56}{Ni} is doubly magic and a Wigner $N=Z$ nucleus, resulting in an increase in $q/A$ compared to the simple mass formula above and the dominance of \nuclei{56}{Ni} in cold, proton-rich NSE. 
 
 \subsection{Free energy for different compositions}
\begin{figure}
\plotone{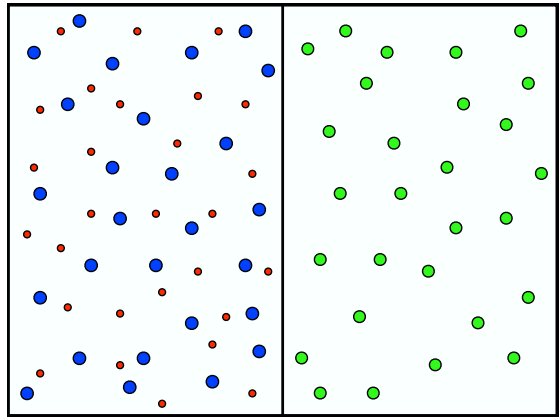}
\caption{\label{fig:ni55_ni56_p} 27 \nuclei{56}{Ni} nuclei and 28 protons on the left, and  28 \nuclei{55}{Ni} nuclei on the right. Both compositions have identical electron fraction $Y_e = 28/55$. Even though pure \nuclei{55}{Ni} has a marginally higher $\bar{Q}$, the NSE state at $\rho\sim10^7 \gcc$ near freeze out is closer to the mix of \nuclei{56}{Ni} nuclei and protons depicted schematically in the left panel.  For almost equal $\bar{Q}$, the composition with more particles and higher entropy is statistically preferred}
\end{figure}

\begin{figure}
\plotone{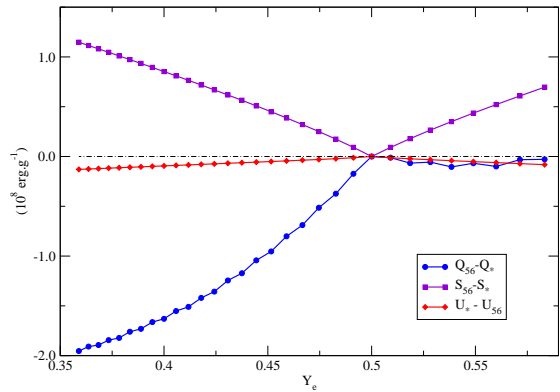}
\caption{\label{fig:diff-ni} Differences in internal energy $U$, entropy contribution $TS$ and binding energy $Q$ between a mix of \nuclei{56}{Ni} (subscript 56) and free protons (for $Y_e>0.5$) or neutrons (for $Y_e<0.5$) and a composition of pure Ni isotope (subscript *) for different \ye.}
\end{figure} 

\begin{figure}
\plotone{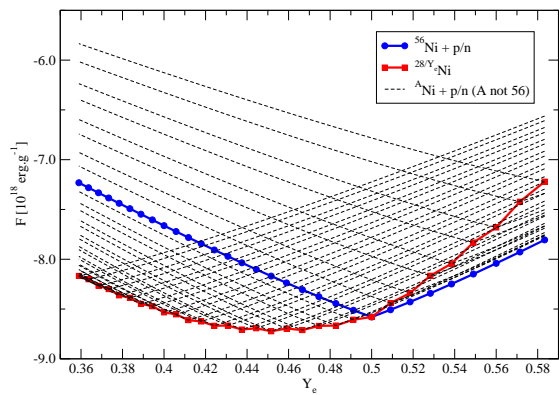}
\caption{\label{fig:ni_Fye} Helmholtz free energy ${\mathcal F}$ of compositions consisting of Ni isotopes and nucleons as a function of $Y_e$. \mbox{$T = 3.5\times10^9\,\mathrm{K}$.} and \mbox{$n_B = 6.0\times10^{30}\,\mathrm{cm}^{-3}$}.  For $Y_e > 0.5$, a mix of protons and test isotope \nuclei{56}{Ni} (thick blue line with dots) is favored. Other test isotopes (dotted lines) result in larger ${\mathcal F}$. For $Y_e<0.5$, a pure isotope composition (thick red line with squares) has the lowest ${\mathcal F}$.}
\end{figure} 

 To get a better understanding for the different low temperature behavior of abundance trends as a function of $\ye$ for the proton and neutron-rich regimes, we consider a simplified system and restrict the composition to nuclides from the Ni isotopic chain and free nucleons. 
 For a concrete example, let us compare two compositions at $Y_e= 28/55 \sim 0.5091$, one consisting of pure \nuclei{55}{Ni}, and the other consisting of a mix of free protons and a ``test" nucleus, here \nuclei{56}{Ni} (see fig.~\ref{fig:ni55_ni56_p}).
The mean binding energy per nucleon is given by $\bar Q = \frac{\sum_i q_in_i}{n_B}$, where $q_i$ the binding energy of nucleus $i$ and $q_n = q_p = 0$. 
 $\bar{Q}$ of 28 \nuclei{55}{Ni} nuclei \mbox{($\frac{28 \times 467.352 \; \mathrm{MeV}}{28\times55} = 8.497$ MeV/nuc)} is slightly larger than that of 27 \nuclei{56}{Ni} nuclei plus 28 protons \mbox{($\frac{27 \times 483.992 \; \mathrm{MeV}}{27\times56\;+\;28\times1} = 8.486$ MeV/nuc).} NSE, however, favors the state with the lowest ${\mathcal F}$, not the lowest $\bar{Q}$. For $T_9=3.5$ and \mbox{$n_B = 6.0\times10^{30}\,\mathrm{cm}^{-3}$} we know that pure \nuclei{55}{Ni} is disfavored and should have higher ${\mathcal F}$. This is indeed the case, as $\Delta U$ between the two states is negligible and the entropy term $TS$ is much larger for the state with the free protons (see fig. \ref{fig:diff-ni}), resulting in lower ${\mathcal F}$. It is evident from fig.~\ref{fig:ni55_ni56_p} that with \nuclei{56}{Ni} as the test nucleus, the increase in the entropy term on the neutron-rich side for the mixed composition is more than compensated by the increase in binding energy by the pure state. On the other hand, for the proton-rich side the $\bar{Q}$ of both compositions is nearly equal, and the higher entropy of the state with more particles makes the difference. 

Using the data from the latest atomic mass evaluation of \citet{audi03} and an equation of state
 \citep{timmes99,timmes00,fryxell00}, we compute and compare ${\mathcal F}$ for such compositions for the whole Ni isotopic chain, with each isotope as the test nucleus at a time.  
If $\ye<Z/A$ of the test nucleus (in the example above and in fig. \ref{fig:diff-ni} this is \nuclei{56}{Ni}), we use free neutrons instead of free protons. 

For any given $Y_e$, we can derive $\bar{Q}$ for a mix of a test nucleus \nuclei{A}{Ni} with the appropriate number of free protons or neutrons,
\begin{eqnarray}
\label{eqn:bind2}
\textrm{if}\ \frac{28}{A} < Y_e,&\quad& \bar Q  = q(\nuclei{A}{Ni}) \frac{1-Y_e}{A-28}  \\
\label{eqn:bind3}
\textrm{if}\ \frac{28}{A} > Y_e,&\quad& \bar Q  = q(\nuclei{A}{Ni}) \frac{ Y_e}{28},  
\end{eqnarray}
 where $q(\nuclei{A}{Ni})$ is the total binding energy of \nuclei{A}{Ni}. 
${\mathcal F}$ of all such compositions is shown as a function of \ye\ in fig.~\ref{fig:ni_Fye}. It  clearly shows that for $Y_e<0.5$ a pure composition of the Ni isotope with $\ye=28/A$ has the lowest ${\mathcal F}$, and that for $Y_e>0.5$ a composition consisting of free protons and \nuclei{56}{Ni} minimizes ${\mathcal F}$. Other choices for the test nucleus have larger ${\mathcal F}$ (dotted lines in  fig.~\ref{fig:ni_Fye}).

\section{Conclusions}
We have presented mass fraction trends for NSE in proton-rich environments.
We have explained the, at first sight peculiar, high free proton mass fractions for $\ye>0.5$ and large mass fraction of \nuclei{56}{Ni} even out to $\ye=0.6$ by considering a simplified model. 
Restricting ourselves to the Ni isotopic chain, we have explicitly shown that 
for $\ye>0.5$ ${\mathcal F}$ is minimized by a state consisting of a mixture of free protons and the self-conjugate \nuclei{56}{Ni}, whereas for $\ye<0.5$ a state with a pure composition made up of the Ni isotope that has $Z/A$ closest to the prescribed $\ye$ is preferred. In reality, other isotopic chains are of course accessible.  Other Fe-peak nuclides (especially the slightly neutron-rich even isotopes of Fe) compete with Ni isotopes for the most tightly bound nucleus with $Z/A$ near $\ye$ (see fig. \ref{fig:lowT}), giving the familiar NSE abundance pattern. 

\acknowledgments
  This work is supported at the University of Chicago by the DoE under Grant B523820 to the ASC/Alliances
  Center for Astrophysical Thermonuclear Flashes, and the NSF under Grant PHY 02-16783 for the 
  Frontier Center ``Joint Institute for Nuclear Astrophysics'' (JINA), and at ANL by
  the U.S. DoE, Office of Nuclear Physics, under contract DE-AC02-06CH11357. EFB is supported by the NSF under grant AST 05-07456.


\clearpage


\end{document}